\documentclass[conference]{IEEEtran}
\ifCLASSINFOpdf
  \usepackage[pdftex]{graphicx}
\else
\fi
\usepackage{subfigure}
\usepackage{url}

\usepackage{comment}
\usepackage{xcolor}

\begin{document}
%
\title{LODUS: A Multi-Level Framework for Simulating Environment and Population - A Contagion Experiment on a Pandemic World }

\author{\IEEEauthorblockN{Gabriel Fonseca Silva, Vinícius Cassol, \\ Amyr Borges Fortes Neto, Andre Antonitsch, \\ Diogo Schaffer, Soraia Raupp Musse}
\IEEEauthorblockA{Virtual Humans Simulation Laboratory\\
Graduate Course in Computer Science\\
School of Technology\\
Pontifical Catholic University of Rio Grande do Sul \\
http://www.inf.pucrs.br/vhlab}
\and
\IEEEauthorblockN{Rodrigo de Marsillac Linn}
\IEEEauthorblockA{Office of the Chief Information Officer \\ Municipal Secretary of Planning and Management \\ Porto Alegre City Hall - RS \\
https://prefeitura.poa.br/smpg} }


%


\maketitle

\begin{abstract}
Nowadays we are experiencing a way of life that never existed before. 
The pandemic has sharply changed our habits, customs, and behavior. In addition, a lot of work was suddenly requested for city managers challenging them to develop strategies to try stopping the pandemic progression. Urban environments must be dynamic and managers need fast decisions when working on crisis situations. In this paper we present LODUS, a framework able to simulate urban environments on a multi-level approach, combining macro and micro simulation information in order to provide accurate information about population dynamics. Furthermore, the framework LODUS is a powerful tool when performing an urban viability study, since the simulation results are able to highlight and predict attention points prior to an urban environment to be built.

\end{abstract}


%
\IEEEpeerreviewmaketitle

\section{Introduction}

\footnote{Draft version made for arXiv: \url{https://arxiv.org/}}A smart city is an urban area that can use electronic sensors to capture information about the population and functioning of spaces. With this information, resources and services can be managed to improve people's quality of life. Besides, the urban ecosystem is constantly evolving and it is important to predict situations which can request immediate actions in order to avoid issues.
The use of technology is a powerful partner when used to help designing, planing and managing a city. In particular, simulations in virtual environments are able to reproduce specific situations to support strategic planning decisions.

Eid and Eldin~\cite{EID1980} discuss that simulation systems can be useful in the simulation of urban areas such as choice of residential locations, problems of transport, etc. In particular, managers are interested in the human behavior associated with the urban planning~\cite{HANDY2002}. Usually, simulation tools are specific to contexts, for instance the mobility study and impact of people's life~\cite{SALLIS2004}. One of the challenges that urban simulation tools present concerns with the level of detail of data and scenarios to be simulated. For instance, if one manager wants to simulate a new route in a public transportation in a city, or inter-state, the data to be included in terms of environment and population changes. So, it is desirable to have a framework where managers can decide the level of detail data that should be present in order to answer the required questions.

This paper aims to present our framework named LODUS, to simulate, with multi level of details, a virtual urban environment taken into account the environment structure and the population who lives on it. We propose a multi-level of detail so the managers are able to see the information about space with the desired detail to take decisions. In addition, we propose and included in LODUS a model to simulate contagion among people in the various levels of details, as a try to help with current situation of COVID-19 pandemic situation.


\section{Related Work}
\label{sec:related}

 In this section we present the related work which supports our research focusing on environment creation and population. In addition, we also explore contagion related research since the population dynamic in the world is currently being affected by the pandemic.

Parish and M\"{u}ller~\cite{parishMuller2001ProceduralModelingOfCities} work presented the first model for procedurally generating different elements of a city, including streets and parcels. The model uses an extension of L-Systems~\cite{lindenmayer1968LSystem} to search for local optimal successors for each rule applied. 
Talton et al.~\cite{talton2011MetropolisProcedural} presented an extension of this model to allow the parameterization of L-System rules and terminal symbols. The model uses a Markov Chain Monte Carlo (MCMC) method to approximate the ideal values for each parameter to create the desired environment. However, this work has presented limited results for larger cities, being limited to a single point of view and requiring a well-defined likelihood function to evaluate each iteration fitness. 

Due to the limiting nature of L-Systems modeling of non-organic elements, different models have been presented to allow higher artistic control following the inverse procedural modeling concept used by Talton et al.~\cite{talton2011MetropolisProcedural}. Vanegas et al.~\cite{vanegas2009interactiveDesign} presented a model for combining behavioral modeling (e.g. population density distribution, job distribution, high access, land usage) and geometric modeling (e.g. road width and number of lanes, building area, number of floors), allowing the changes made in a certain parameter to impact on every element of the city layout. Vanegas et al. ~\cite{vanegas2012inverseDesign} presented a model where, given an initial city model, adjustments using an MCMC method are made to approximate a set of parameters (e.g. parcel average area, road curvature, building height) defined by the user. The model presented by Aliaga et al. ~\cite{aliaga2017FastWeatherSimulation} allows the user to draw the distribution on land usage (e.g. urban regions, agricultural regions, bodies of water) on the base terrain to create weather simulations. Roads and buildings for urban regions are adjusted using a Metropolis-Hastings algorithm~\cite{metropolis1953, hastings1970MonteCS}, an MCMC method, to explore the search space and find similar results to the user-specified weather. Specifications can include cloud coverage per region, humidity, rain distribution, and city temperature. Mustafa et al.~\cite{mustafa2018floodSensitive} presented a model for the generation of urban layouts that passively reduce water depth during flood scenarios. This work combines a procedural generation model with a hydraulic model to evaluate water flow characteristics for the created region. A neural network is used to identify relationships between the urban layout and the flow of water during floods, which are used as input to an MCMC model, adjusting the environment. The evaluation takes into consideration the desired building coverage and average water depth, etc.

Populations and crowds can be simulated with varying levels-of-detail, with a trade-off between accuracy and computational performance. While a microscopic crowd simulation offers individual characteristics and decision making for each agent, it can be prohibitively computationally demanding when simulating an environment such as an entire city. Macroscopic crowd simulation models group up agents to reduce the granularity of the simulation and simulate larger crowds. One of such models is BioClouds~\cite{antonitsch2019bioclouds2}, a crowd simulation model which offers collision avoidance based on space discretization and competition. BioClouds models crowds as \textit{clouds}, i.e., groups of similar minded agents, which have a desire for a certain density, speed and goal. Clouds compete for space amongst each other, and occupy the environment in a manner that tries to keep their desired densities respected. 

If on one hand, BioClouds simulate macroscopically groups, sometimes the microscopic simulation of people is needed. The crisis caused by the new Corona Virus and the global spread of COVID-19 cases turned the attention of scientific community to the spreading disease problem. Some research groups are focusing on SIR~\cite{kermack1927contribution} (further formulated in Section~\ref{sec:populationAndContagion}) and SEIR~\cite{engbert2020sequential} mathematical models to attempt to predict when the flatten of contagion curves will occur. It is of major importance to predict those curves, given every scenario~\cite{JTD36385}. Authorities must be able to decide when to open schools, stores and shopping centers with the objective to protect children education and jobs, 
minimizing the risk of 
endangering public health. 

In this paper we use the model proposed by Antonitsch~\cite{antonitsch2019bioclouds2}, where we do not simulate individuals, but groups, in a macroscopic level.  Therefore, we also propose to include in LODUS a microscopic simulation using BioCrowds in order to have individuals. Our goal is to customize BioCrowds to program agents to attempt to keep recommended social distance. Then, we extract information about social distancing simulation to estimate contagion rates for micro-environments, then we extrapolate those data for macroscopic simulation. 

\section{LODUS: Level-of-Detail on Urban Simulation}

\begin{figure} [htb]
    \centering
    \includegraphics[scale = 0.6]{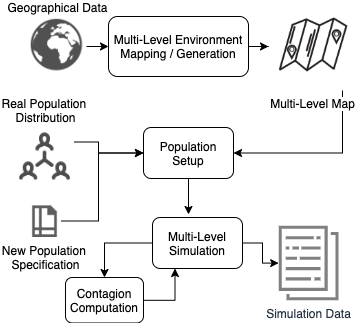}
    \caption{LODUS Architecture.}
    \label{fig:model}
\end{figure}

The ability to predict and analyze urban dynamics scenarios is a key and distinctive support for the work of city managers and urban planners. This paper presents LODUS, a framework able to simulate virtual urban environments with various levels of details. LODUS main goals is to provide information in order to collaborate on the challenge of city planning and management. Figure~\ref{fig:model} illustrates the architecture of the proposed framework. 

LODUS is a multi-level model able to be configured in two ways: i) reproducing scenarios of real world or ii) simulating urban dynamics before the introduction of changes in real life. In both cases, it is important that environment and population could be coherently simulated. The next sections describe in details every module which compose our multi-level framework.

\subsection{Environment}

\begin{figure*}[htb]
  \centering
  \subfigure[fig:pipelineBoundaries][Environment Bounrdaries]{\includegraphics[width=0.23\textwidth]{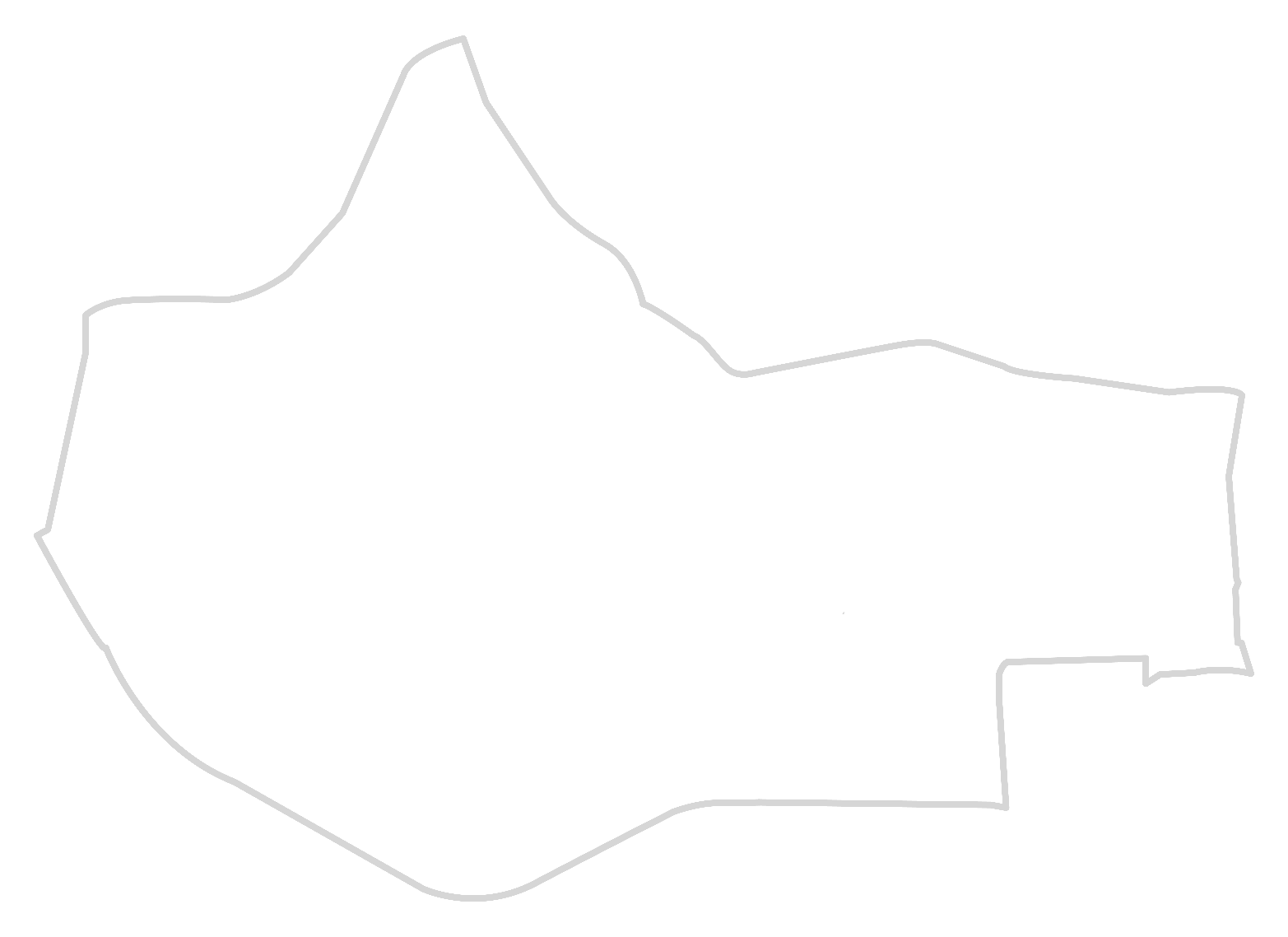}\label{fig:environmentBoundries}}
  \subfigure[fig:pipelineA][Region Graph]{\includegraphics[width=0.23\textwidth]{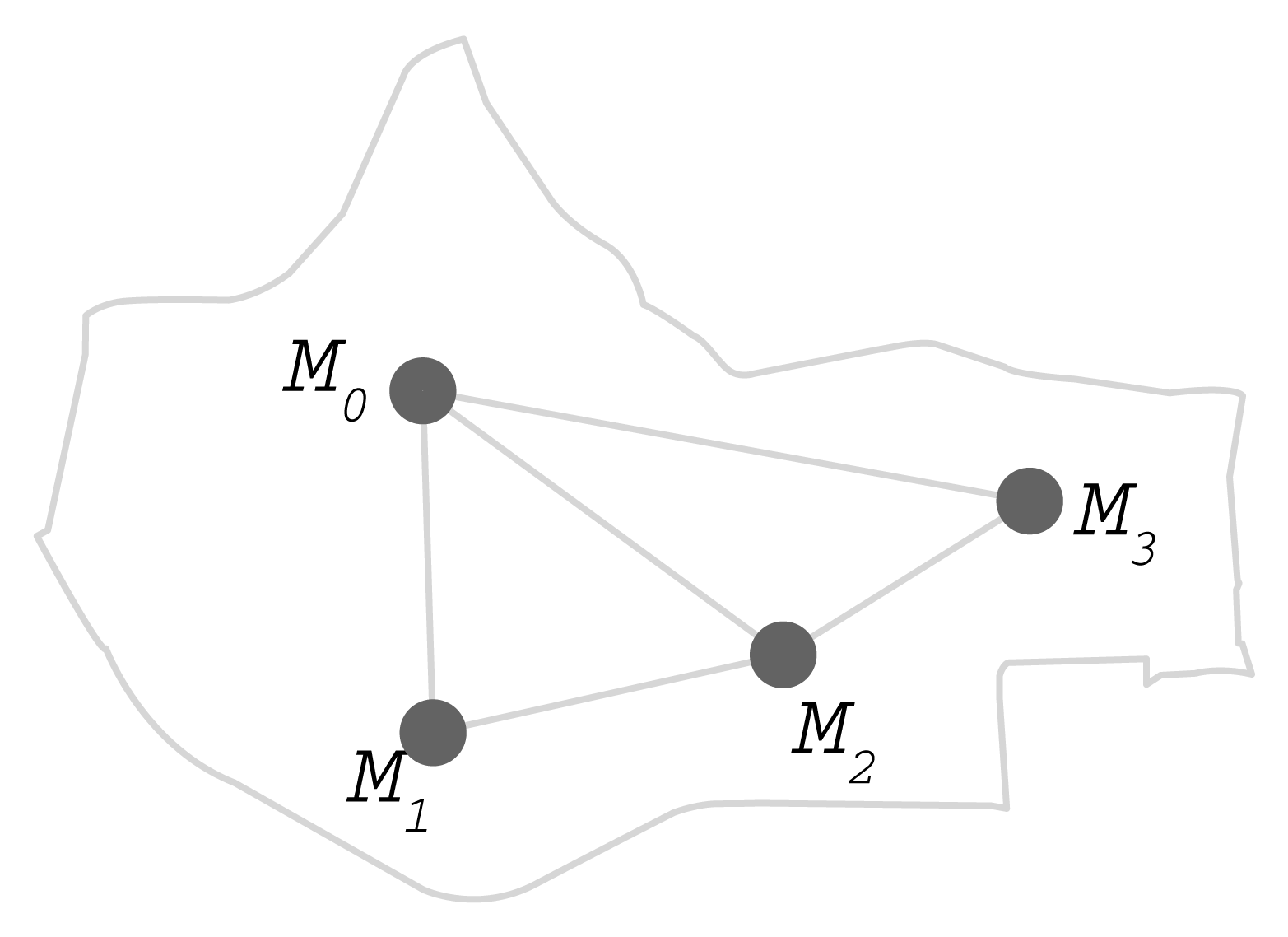}\label{fig:regionGraph}}
  \subfigure[fig:pipelineB][Region Outlines and Connections]{\includegraphics[width=0.23\textwidth]{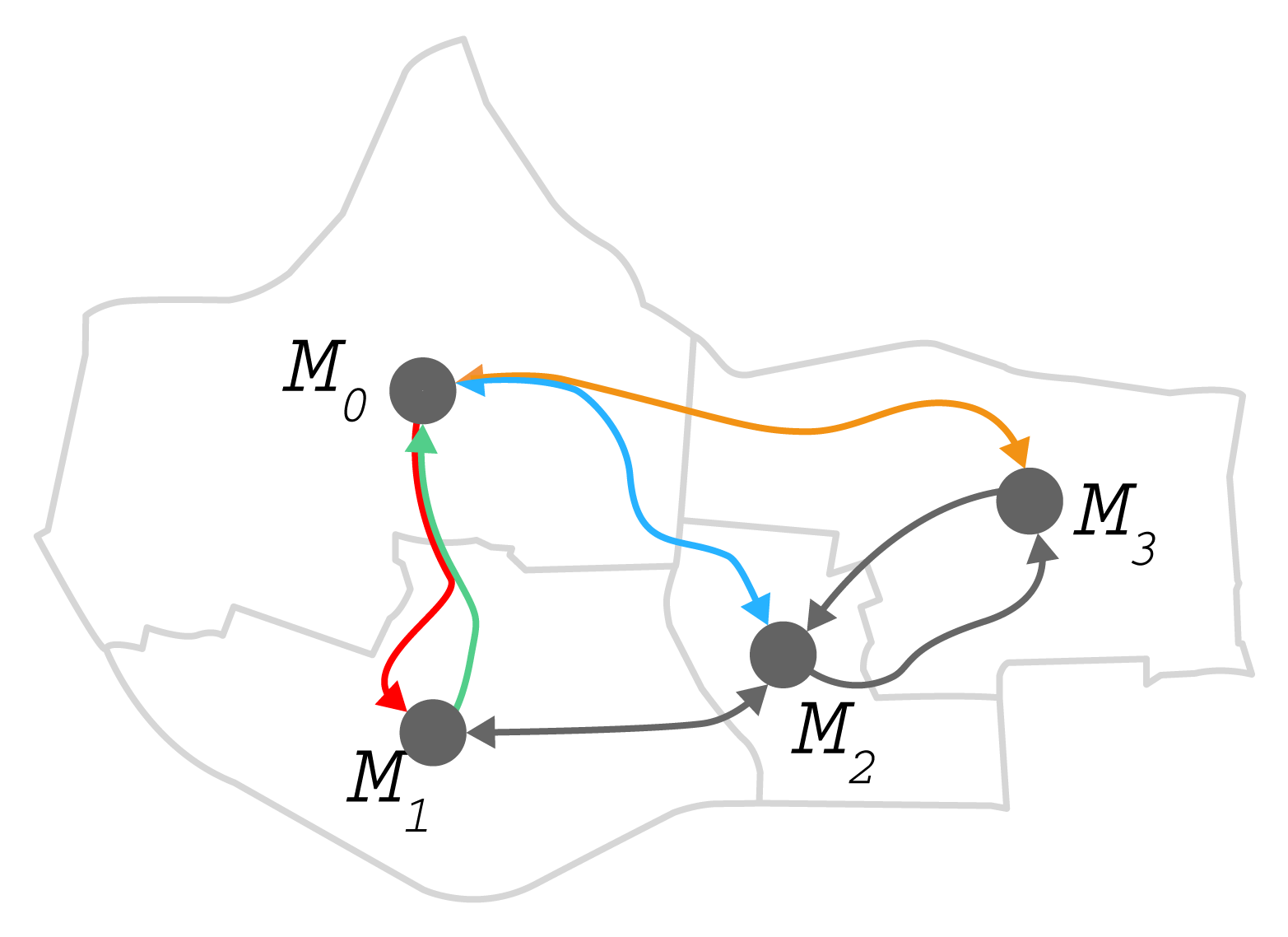}\label{fig:regionOutlines}}
  \subfigure[fig:pipelineC][Division into Subregions.]{\includegraphics[width=0.23\textwidth]{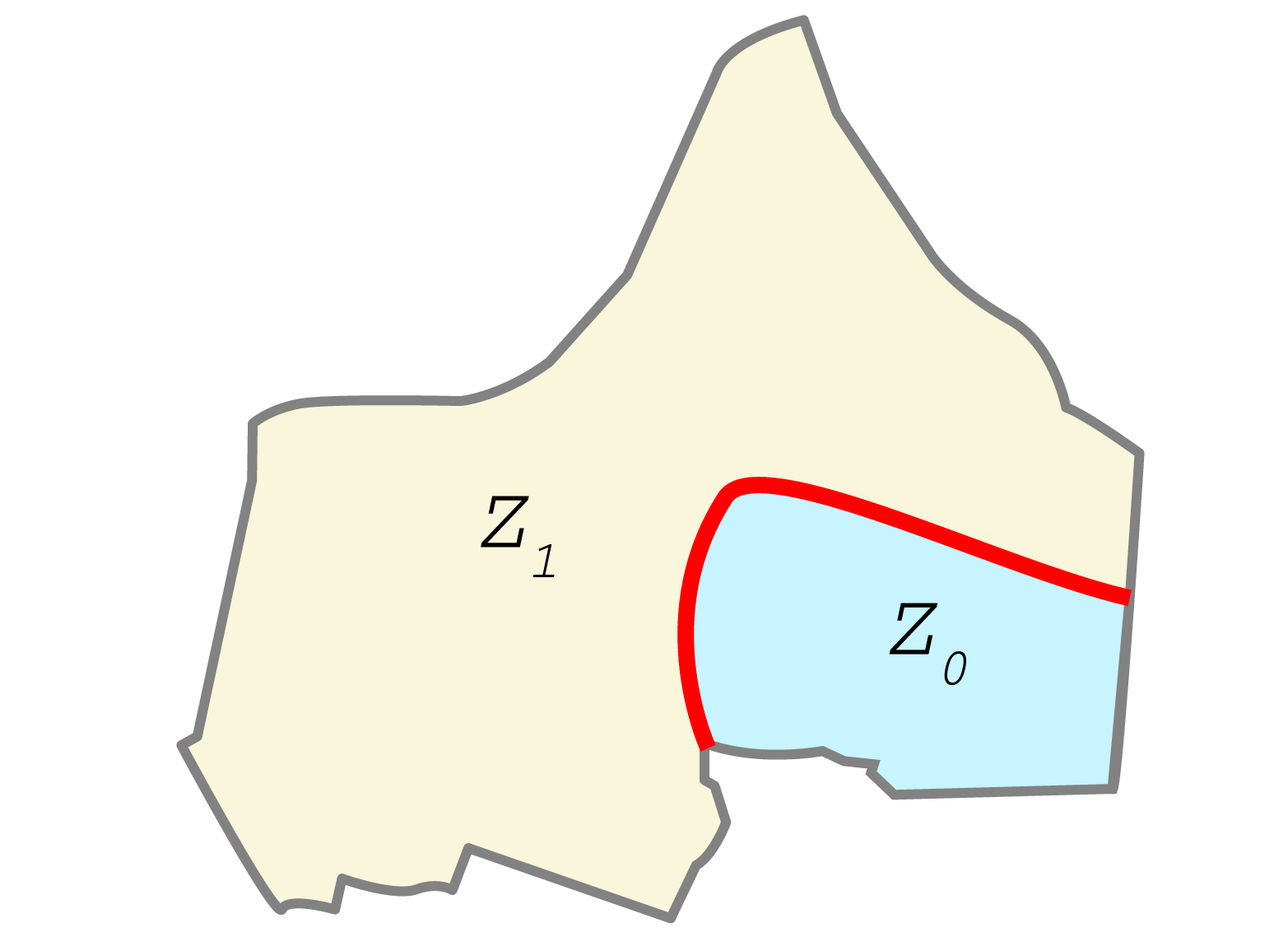}\label{fig:subregionDiv}}
  \subfigure[fig:pipelineD][Definition of a Road Network]{\includegraphics[width=0.23\textwidth]{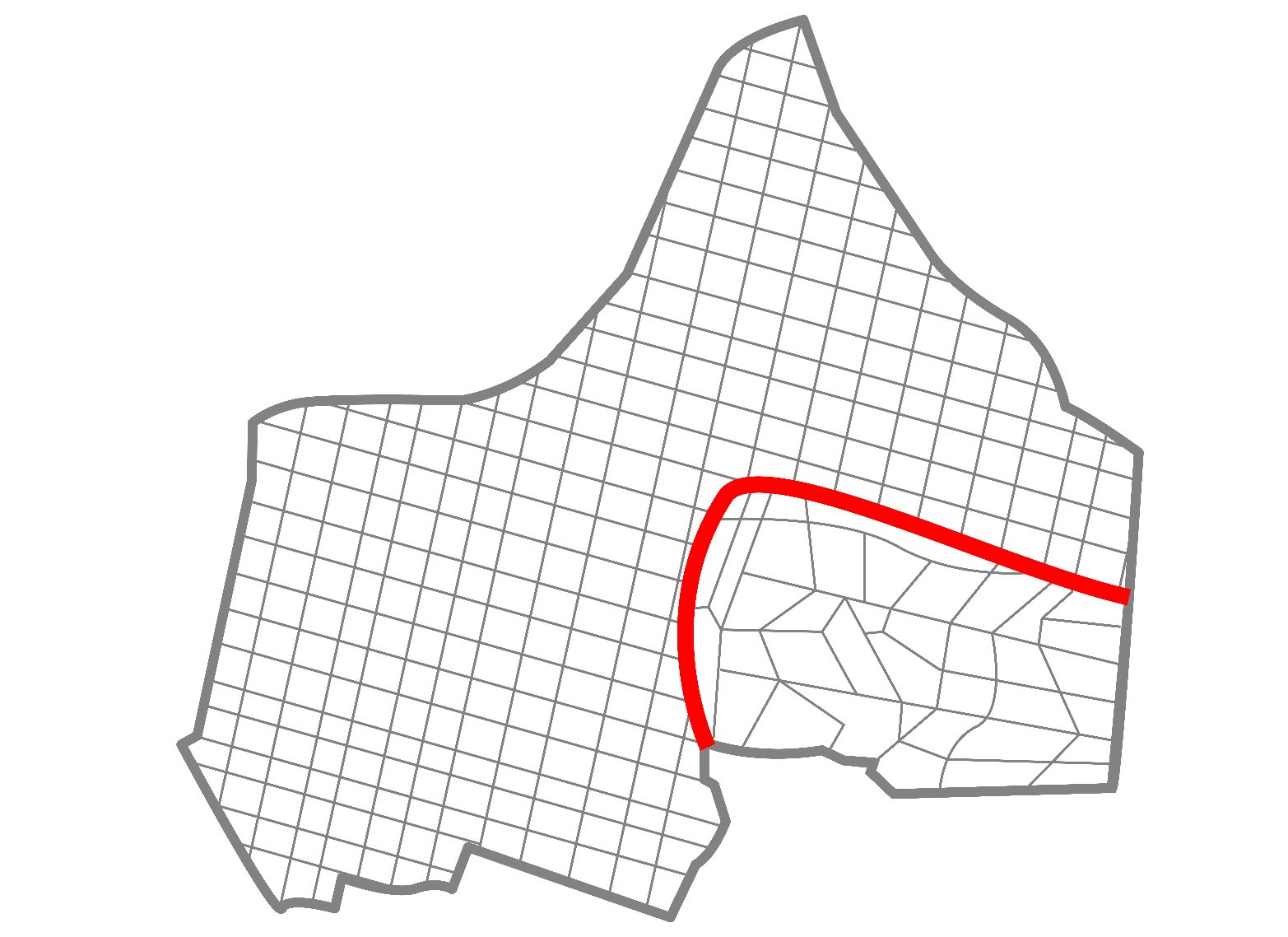}\label{fig:roadDefinition}}
  \subfigure[fig:pipelineE][Identification of City Blocks.]{\includegraphics[width=0.23\textwidth]{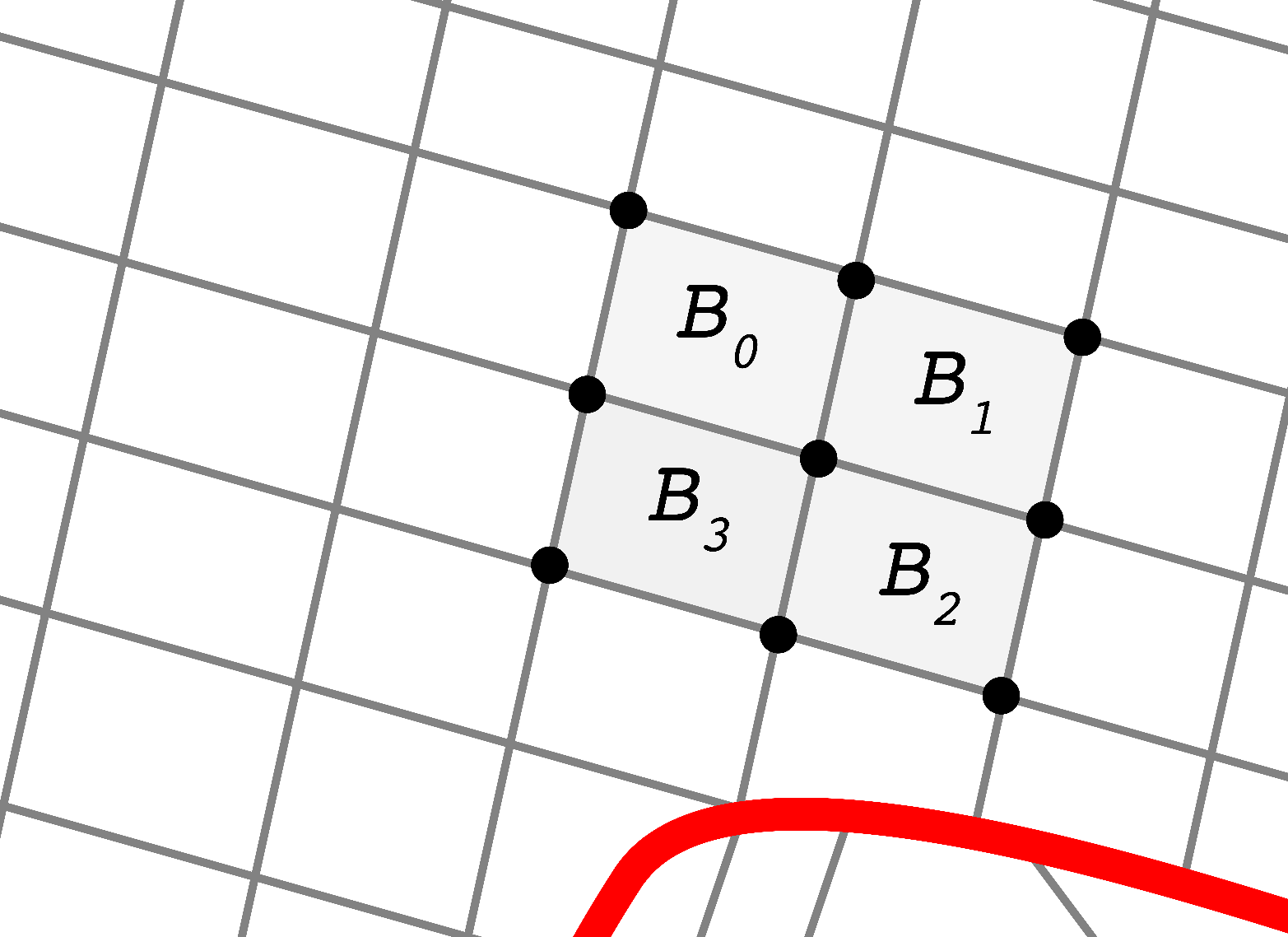}\label{fig:cityBlocks}}
  \subfigure[fig:pipelineF][Parcel Subdivision]{\includegraphics[width=0.23\textwidth]{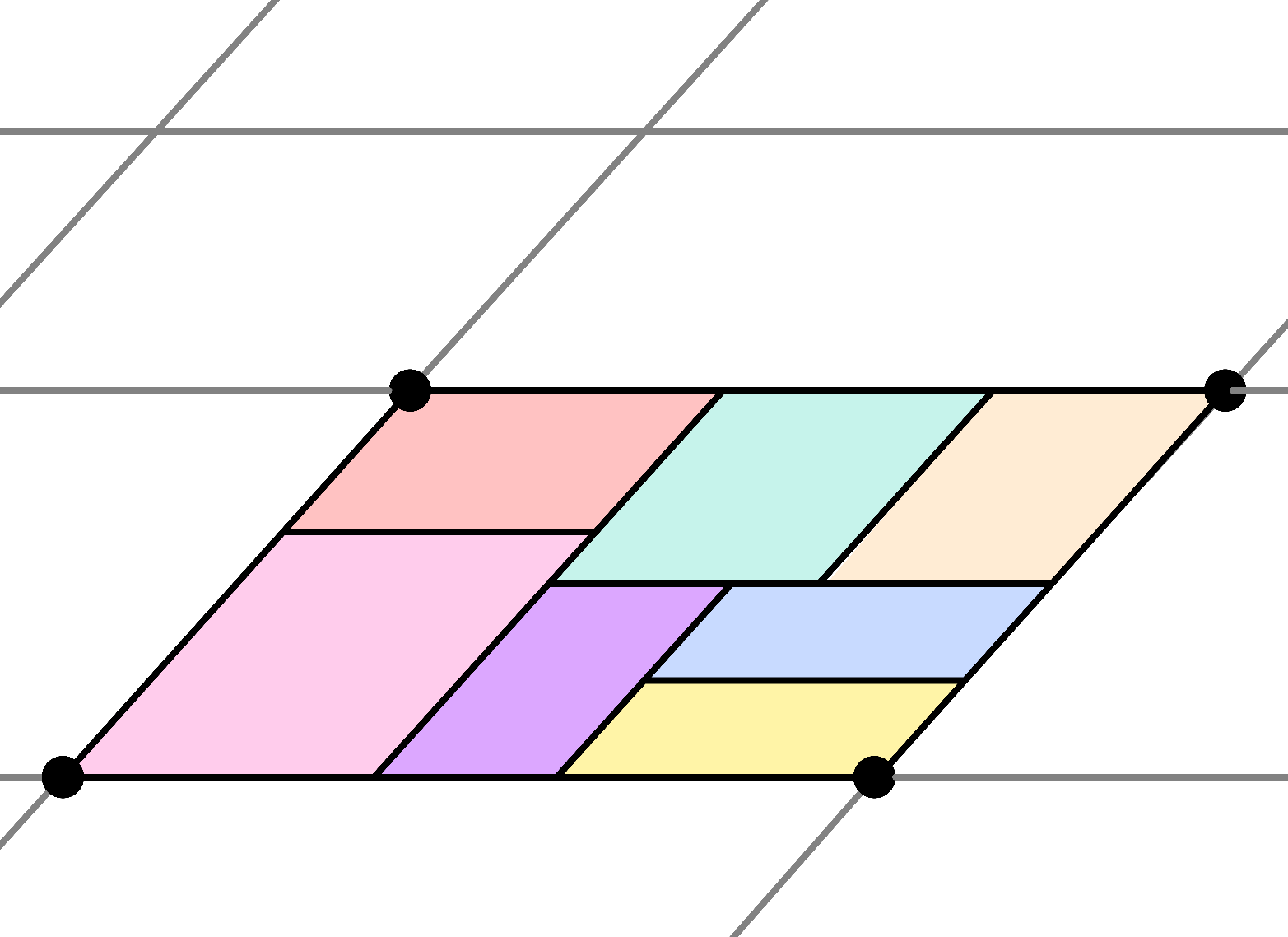}\label{fig:parcelSubdiv}}
  \subfigure[fig:pipelineG][Placement of Buildings]{\includegraphics[width=0.23\textwidth]{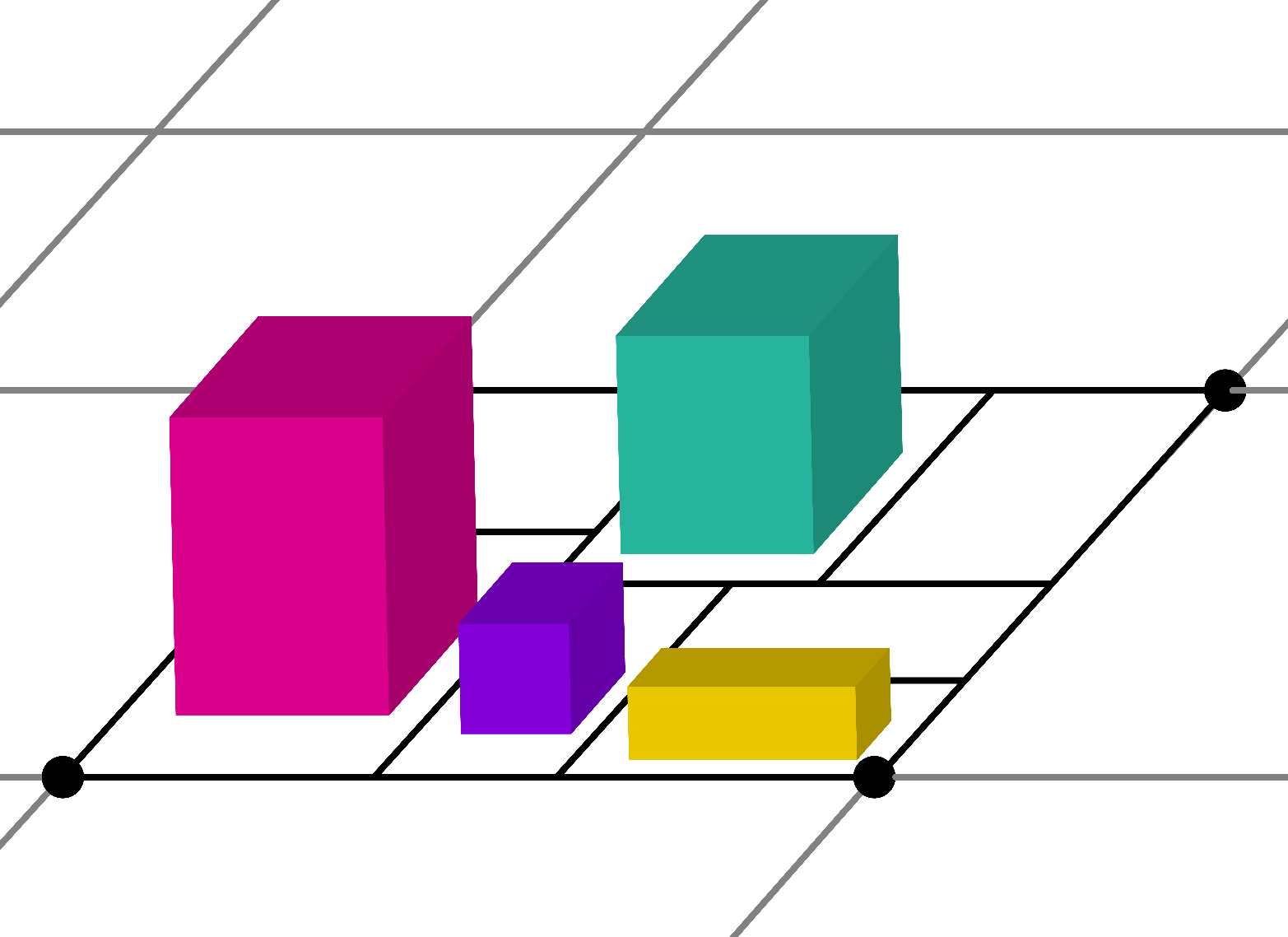}\label{fig:buildingPlacement}}

    \caption{Overview of the framework's stages.}
    \label{fig:Environment}
\end{figure*}

This module allows the representation of an environment with different levels of abstractions. At each level, a more detailed representation of the environment may be defined. As a deeper level is included, more detailed information can be computed. Such details store different data, such as population distribution, road system, buildings' geometry or even internal buildings setups. Figure~\ref{fig:Environment} presents the steps performed on the environment construction.

First, the \textbf{Environment Boundaries} are represented in a 2-D plane as illustrated in Figure~\ref{fig:environmentBoundries}.
We can define the environment as a set of \textbf{Regions}, where each Region($M$) is a 2-D point placed within the environment's bounds. A Region can represent different areas according to the level of abstraction desired, such as:
\begin{itemize}
\item City block groups within a neighborhood.
\item Neighbourhoods within a single city;
\item Cities within a metropolitan area;
\item Counties within a state; and
\item States within a country.
\end{itemize}

The \textbf{Region Graph} (Fig.~\ref{fig:regionGraph}) can be specified as an undirected graph $G = (P,E)$, where $P$ is a set of nodes (Regions) and $E$ contains a set of edges connecting the nodes.  An edge is a straight line segment between a pair of points with weight equal to its euclidean distance. A graph embedding must be guaranteed so there are no inconsistencies between the connections. Without that property, it is possible to establish a road that overpasses an entire Region, ignoring intersections and buildings. Since a planar graph can be nested within another graph's node, it is possible to define different levels of abstractions for the Region Graph. 

To allow the creation of a road network and geometry of buildings, each region $M$ is also represented by an \textbf{Outline} (Fig.~\ref{fig:regionOutlines}). The outline of a region ($O_M$) is a polygon described by a set of vertex that represents its geometry. 
Similar to the Region Graph, the set of outlines must not contain overlapping sections to ensure the consistency of road placements. This can be achieved by allowing each outline point to belong to multiple neighboring regions.


Following, a connection represents a route that allows movement from $M_a$ to $M_b$ through an ordered list of 2-D points in space, where every point is within the geometry of $O_a \cup O_b$. Since a road network may contain one-way streets, connections between regions are directional, but they can be used in reverse if desired by the user and permitted by the constraints of the environment (i.e two-way street representation). Segments of a Connection may also be shared between different regions to allow movement in similar routes. 


To represent the urban layout in a lower level of abstraction, it's necessary to establish a \textbf{Road Network} (Fig.~\ref{fig:roadDefinition}) within each Region of the environment. A Road ($Q$) is defined by a list of 2-D points in space, with each pair of points ($[Q_{[i]}, Q_{[i+1]}]$) representing a road segment. A set of points in a road can be shared with an Outline, allowing the road to represent the limits of a Region. To preserve the properties of a planar graph inherited by the Region Graph, road segments may not cross each other. In cases of intersections, a new point is placed in the environment and the list on each road is adjusted to comprise the new segment. 
Even though they share the same structure, the framework distinguishes roads into ``primary" (i.e. arterial roads, main avenues or highways) and ``secondary" (i.e. local streets), due to their purposes at each step. 

Primary roads cross large portions of a Region outline and are responsible for dividing the outline into \textbf{Subregions} (Fig.~\ref{fig:subregionDiv}) representing a segmentation of an Outline. Secondary roads provide local access for different areas inside a Subregion and can have a smaller length. The extremity points of a primary road must belong to either the Region outline or another primary road. This constraint is applied to maintain the planar graph properties. 
A set of parameters is defined to describe roads in higher detail and allow integration with other applications (e.g traffic simulation). 

\begin{itemize}
\item $Width$: Total width of the road;
\item $Lanes$: Number of lanes in the road;
\item $IsOneWay$: Indicates if a road is one-way. If true, the ascending order of the point list indicates the traffic direction.;
\item $MaxSpeed$: Max driving speed allowed for vehicles in the road; and
\item $MaxHeight$ and $MaxWeight$: Max height and weight allowed in the road, be that for legal reasons or physical limitation (e.g. an overpass in a nearby intersection, high variation in elevation, or crossings over bridges or unstable terrain).
\end{itemize}


Because the planar properties of the Region Graph are preserved through each step of this multi-level process, Subregions can be identified using a cycle detection algorithm for undirected graphs. 


Conceptually, each Subregion represents an area that shares a profile of buildings, road infrastructure, and population distribution. For example, a Subregion that represents a suburban area may have a grid-like road network, with the predominance of smaller houses over tall buildings and a higher number of parks; a city center may have a higher number of apartment buildings, therefore a higher population density. 

Once the network of secondary roads is formed, it's possible to obtain a set of \textbf{City Blocks} (Fig.~\ref{fig:cityBlocks}) within each Subregion. A $Block$ is also represented by a set of 2-D points forming a polygon. Each point in a City Block belongs to either a secondary road segment or the Subregion outline. 
City blocks comprise a set of \textbf{Parcels} (Fig.~\ref{fig:parcelSubdiv}), which are used as a base for the generation of \textbf{buildings and houses} (Fig.~\ref{fig:buildingPlacement}). Blocks within the same Subregion may share a default set of properties to give them a similar look.
Since a City Block can contain multiple land Parcels, a subdivision algorithm is applied to the desired blocks with a large enough area. The process is based on the work of Parish and M\"{u}ller~\cite{parishMuller2001ProceduralModelingOfCities}, and Vanegas~\cite{vanegas2012parcelGeneration}, where a series of splits are applied to a geometry according to the following steps:

\begin{enumerate}
    \item the larger axis of an Oriented Bounding Box (OBB) is identified.
    \item a cut is made perpendicularly across the axis mid-point.
    \item this process is repeated recursively until the area of a new Parcel is below a specified threshold. 
\end{enumerate}

In this method, land Parcels are approximations of quadrilaterals, with exceptions being in cases where the Parcel makes contact with an irregular road (e.g. rounded corners or dead ends). An offset value (in \%) can be used to change the axis mid-point at each iteration, allowing a larger variety of outputs. 
According to this process, a cut can result in a Parcel being created inside the geometry without street access. Following user preferences, different solutions can be applied in these scenarios: 
\begin{enumerate}
    \item the Parcel is flagged as "non-building", being used to generate parks, playgrounds and empty lots;
    \item the cut is reverted and the previous Parcel is validated, ending the recursion;
    \item the cut is reverted and the smaller axis is selected for the split, creating two narrower lots, with both having street access; 
    \item a secondary street is created following one of the OBB's axis until the lot is reached; and
    \item  a local access (i.e alleyway) is created, not altering the City Block geometry.
\end{enumerate}

It is important to mention that the procedural generation of buildings geometry is out of the scope of this paper. However, many methods found in the literature use a 2-D polygon as input for their process, defining a foundation for the architecture. To allow the application of these grammar-based methods within our model, we propose a set of parameters that can produce the results desired by the user. Primarily, a building footprint must be defined to serve as usable areas. This footprint can be obtained as the outline of each valid Parcel generated by the recursive cuts. For models requiring a rectangular area as input~\cite{rodrigues2010dataModelText, greuter2003pseudoInfiniteCities}, the largest inscribed rectangle within the outline can be used. Since Parcels facing only one road segment have well-defined side edges, an inscribed rectangle using these edges will align the orientation of the building to face the segment.

Due to neighboring Parcels sharing edges, it's possible that their footprints also share a segment, causing the generated buildings to be very near to each other. A set of setbacks ($Setback_{front}$ and $Setback_{side}$) are applied to reduce the usable area of each Parcel, creating a space for sidewalks, vegetation, and environment props. $BuildingHeight$ (i.e. number of floors) is randomly selected from a range, specified for each City Block, or shared between every block in a Subregion.

In this final stage of the \textit{Environment Generation Process}, a high-detail representation of an urban layout is presented, allowing it's application in fields focused on microscopic elements of a city. Parcel splitting and building parameters for a Subregion can be used to standardize new suburban areas in a growing city.
Also, the setback parameter applied to Parcels allows for the representation of sidewalks in each City Block. Crosswalks between City Blocks can be generated based on the profile of the adjacent road segment. For example, segments belonging to primary roads are more sparse and require the placement of traffic lights; while segments of secondary roads are 
more restricted. This "zoomed in" representation of a Subregion can be independently used as an input for smart cities approaches as well as in the fields of crowd simulation, virtual reality, and digital games.
When the different levels for the environment are defined, the framework is able to produce a multi-level map which is considered to integrating coherently population dynamics in the urban environment. Next section details such integration.

\subsection{Population and Contagion}
\label{sec:populationAndContagion}

We propose using BioClouds~\cite{antonitsch2019bioclouds2} to simulate population movement on a city-sized scale. Clouds in BioClouds represent groups of individuals which move towards a common destination, and sharing some common desired characteristics, such as: desired speed, desired density, and goal. It is used to simulate the overall large-scale population movement in the simulated city (macroscopic), for example, the movement of people between neighborhoods, such as seen in Figure~\ref{fig:regionOutlines}. The movement of clouds between regions lets us define an average occupation of each neighbourhood at any given moment during a simulation. In BioClouds, the user can define the behavior of a cloud by setting their goals, desired radius, desired speed and population quantity. In this framework, we propose creating clouds with information that the user is interested in modeling, e.g. time of movement, population quantity, trajectories and etc. The BioClouds environment is defined by a regular grid which represents markers agents can compete for, each of these markers is a square with area defined by the user, to model environments of different sizes.

We add to each cloud a set of population related characteristics, so that each cloud represents a certain profile of population. This additional set of characteristics is also used to track infection in groups, which is computed in a microscopic simulation, based on the macroscopically-computed neighborhood occupation and densities. This separates the overall flow of population in a city and the local individual movement of agents in a local space.

The contagion is promoted based on SIR dynamics. A subset of the population serves as input for the contagion model, which in turn performs a contagion event. In every event, a number of agents move from Susceptible state to Infected, and from Infected to Removed, depending on: i) information about the group regarding the number of initial infected agents and the group size, ii) the average distance between group members, represented by its density, iii) a threshold defined by the analysis of data acquired in simulation scenarios, and iv) a random component. We now explain each one of these dependencies. 

The SIR model~\cite{kermack1927contribution} considers that agents belong to one of three states: Susceptible, Infected or Removed. The last one stands for people who can not impact contagion anymore, either by acquiring immunity (recovered), or by death. The numeric dynamic of the SIR approach is defined in Equations~\ref{eq:deltaS}, \ref{eq:deltaI}, and \ref{eq:deltaR}.

\begin{equation}
\label{eq:deltaS}
   { {dS}\over{dt} } =- { {\beta I S}\over{N} },
\end{equation}

\begin{equation}
\label{eq:deltaI}
   { {dI}\over{dt} } = { {\beta I S}\over{N} } - {\gamma I},
\end{equation}

\begin{equation}
\label{eq:deltaR}
   { {dR}\over{dt} } = {\gamma I},
\end{equation}
where, ${dS}\over{dt}$, ${dI}\over{dt}$ and ${dR}\over{dt}$ are the variation over time of Susceptible, Infected and Recovered in a population of $N$ individuals. The parameters $\beta$ and $\gamma$ represents the rate in which people become infected and removed respectively. The basic reproduction ratio, denoted as $R_0$, is defined as $R_0 = { \beta \over \gamma}$, and for COVID-19, it has been defined as $R_0 = 2$. Values of $\beta = 0.7$ and $\gamma=0.35$ satisfy this condition and are used herein as parameters for the SIR model. One important property of the SIR approach is that $N$ is constant, and the equality:

\begin{equation}
\label{eq:propertySIR}
   { S(t) + I(t) + R(t) } = {N},
\end{equation}
remains true all the time.

During the simulation of city dynamics, contagion events are triggered by BioClouds, and a subset of the population is submitted to contagion. A number $N$ of agents belonging to the subgroup is the input of our contagion model, along with the information of state on each population member (S, I or R), and a density dependant on the area where the subgroup is placed. This can be anything, from a city, to a district or even a building or public transportation system.

Depending on the density, we estimate the chance of an infection occurring in given circumstances. Simulations were performed varying the number of agents, in order to evaluate how long people violate recommended social distancing. The violation of social distancing occurs when one agent stay inside any other agent's isolation space. For this work, we set this distance to be 2 meters. With the data collected during simulations, we generate a graph correlating the violation rate and density. The result is the ratio of social distancing violation, which reflects on the infection hazard and is used as threshold $T$ (in the range $[0, 1]$) to decide whether to spread infection. At this point, we introduce a random component $r$, also in the range $[0, 1]$ and if $r \leq T$ then the contagion occurs, otherwise the event is ignored and the group's SIR state remains unchanged.

\begin{figure}
    \centering
    \includegraphics[scale = 0.27]{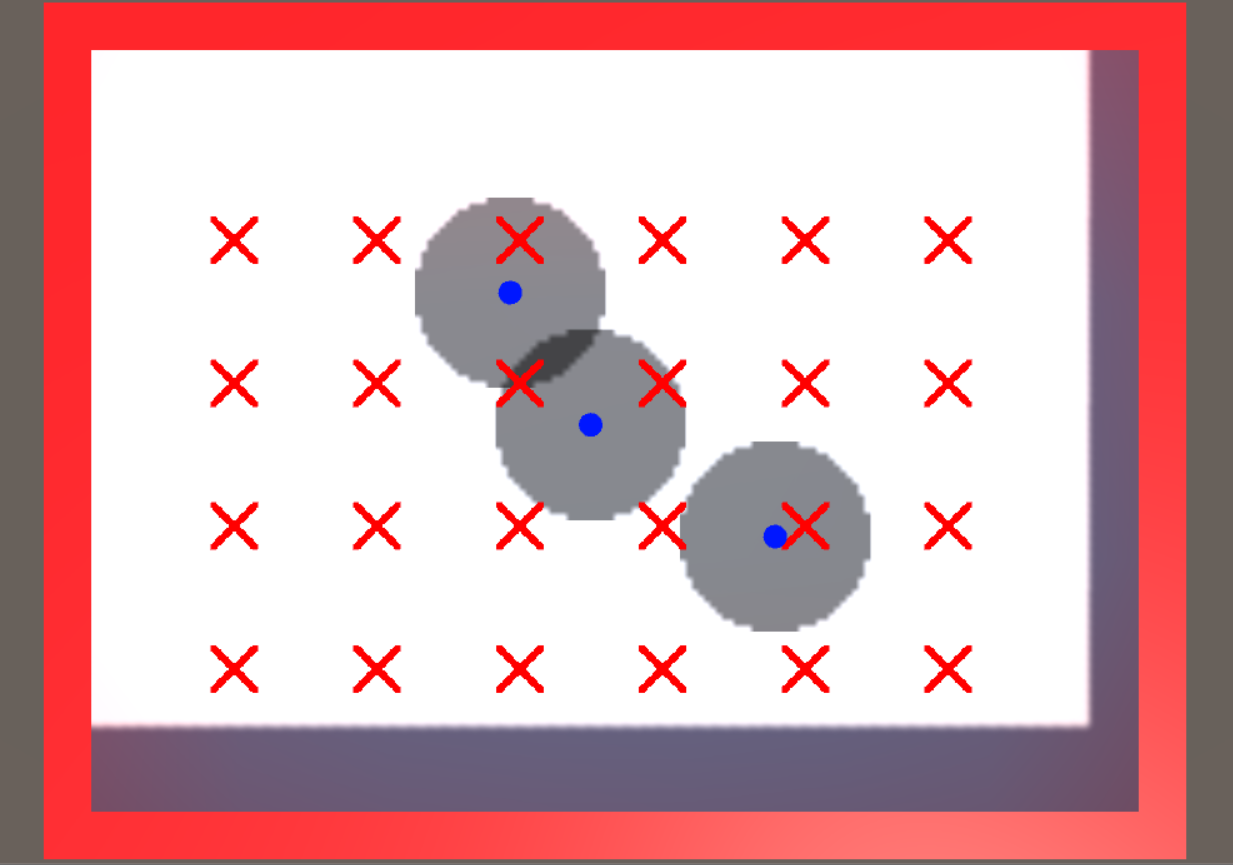}
    \caption{Scenario used to collect data on social distance violation ratio as a function of agent density.}
    \label{fig:simulationScenario}
\end{figure}

Finally, to illustrate the process of determining the social distancing violation ratio, we simulated a scenario with dimensions $22m \times 16m$ resulting in an area of $352m^2$, as depicted in Figure~\ref{fig:simulationScenario}. In the figure, one can see three agents, depicted as blue circles, and their respective social distancing space, depicted as transparent grey circles around them. The red X-marks are goals for the agents, chosen so agents move randomly around the environment. When an agent reaches her/his goal, it randomly chooses another one, using uniform distribution. The scenario is delimited by red walls, and represents indoor environments. The simulations run for two minutes, during which the minimum distance from each agent to others is measured.

While agents move randomly in the scene, they are not supposed to enter each other social distancing space. But, agents have a priority given by their position in the array, and sometimes lower priority agents get stuck between higher priority agents. At this point, social distancing may be violated, as an emergent behaviour inherent to the underlying space competition nature of BioCrowds~\cite{bicho2009modelagem}, which is the microscopic approach on BioClouds, and thus configuring an infection hazard condition. For every frame this condition is detected, a counter is incremented, and by the end we compute the ratio of infection by dividing this counter with the total number of frames in the simulation. In the best case, no violation would be detected, leading to a ratio of zero. In the worst case, agents will violate social distancing in every frame, resulting in a ratio of one.

\section{Obtained Results}

This section presents the real application of LODUS to model a real urban scenario, applying different levels of abstractions with increasing information to reproduce coherent environment and population\footnote{Complementary video material available at: \url{https://youtu.be/TYCvGEONbyg}}.
To start the experiment on a real city, we collected real information to be considered as input for our multi-level approach for environment and simulation.  Such information was gathered from the following source:

\begin{enumerate}
    \item Geographic information from OpenStreetMap\footnote{OpenStreetMap is available at \url{https://www.openstreetmap.org/}} (OSM) dataset, a collaborative mapping project; 
    \item Information regarding the city's neighborhood and population distribution was obtained via the city hall website (http://datapoa.com.br/group/meio-ambiente);
    \item Geographic information system application QGIS\footnote{QGIS is available at \url{https://www.qgis.org/}} was used to align information of different sources within the same coordinate system; and 
    \item Unity3D\footnote{Unity3D is available at \url{https://unity3d.com/}}  was applied to produce a 3-D visualization layer. 
\end{enumerate} 

\subsection{Environment Reproduction}
To represent the city in a higher level of abstraction, a Region Graph was constructed utilizing an entire neighborhood as a base unit for a Region within the graph. Therefore, the graph represents the connections between close neighborhoods.

\begin{figure}[!htb]
  \centering
  \subfigure[fig:regionGraph][Region Graph.]{\includegraphics[width=0.18\textwidth]{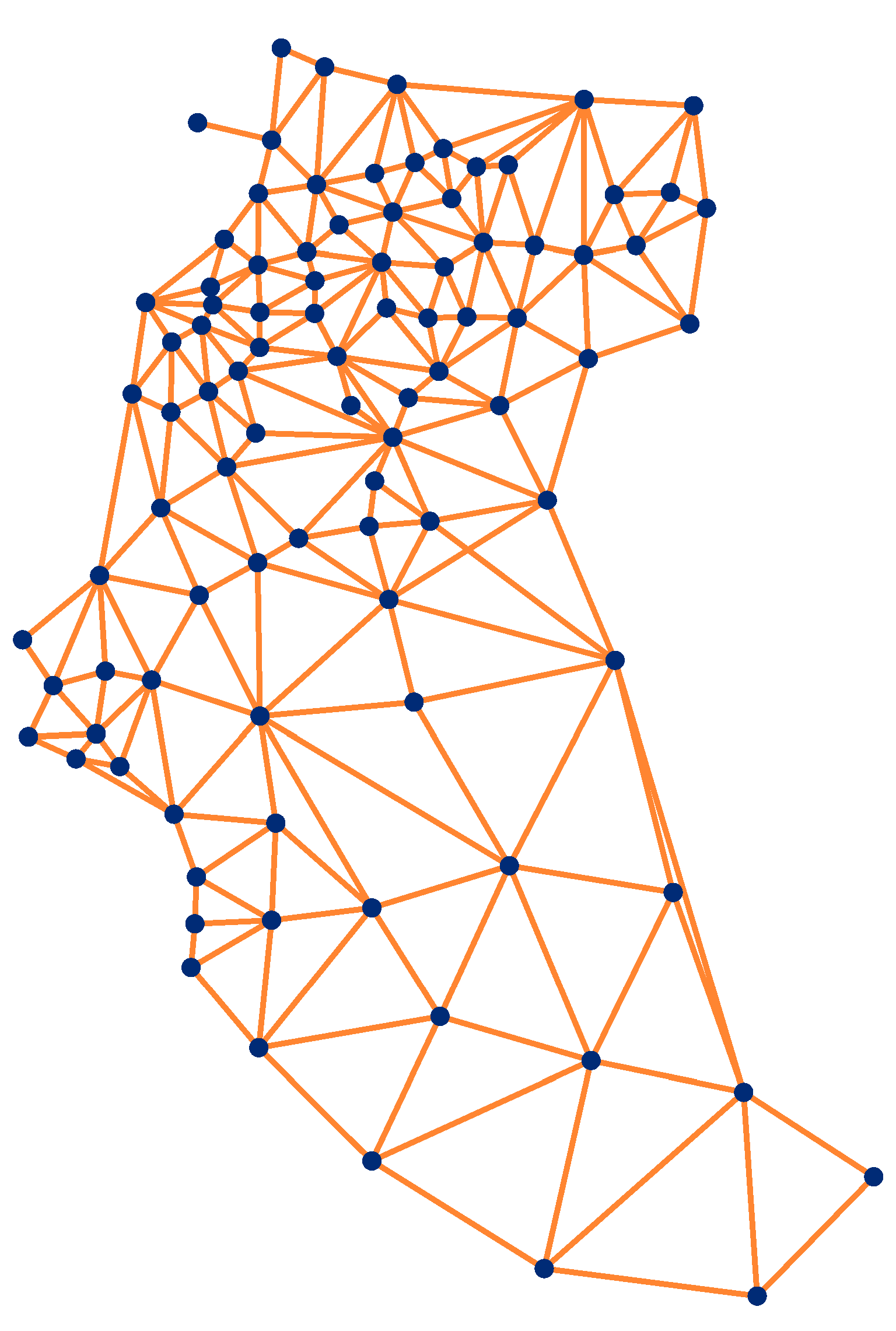}\label{fig:regionGraphFigure}}
  \subfigure[fig:regionOutlines][Region Outlines.]{\includegraphics[width=0.20\textwidth]{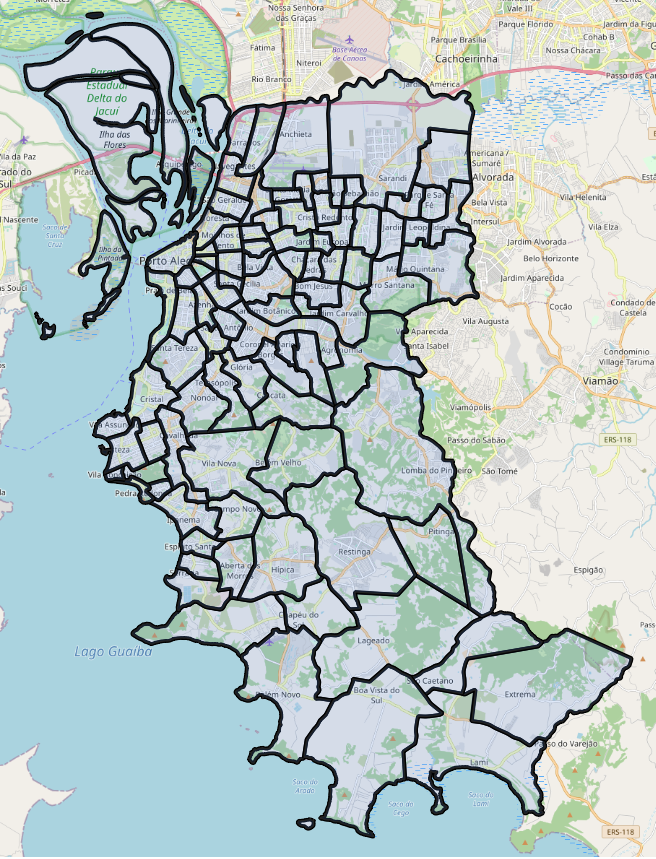}\label{fig:regionOutlineFigure}}
    \caption{Region Graph and Outlines.}
    \label{fig:graphAndOutline}
\end{figure}

The resulting \textbf{Region Graph} is presented in Figure~\ref{fig:regionGraphFigure}. An edge between two Regions was defined for every case were two neighborhoods share a boundary, be that a physical road or an imaginary line. 
To allow a closer representation of the real-world city and an easier integration with the OSM dataset, the Region Graph was constructed using the decimal Latitude-Longitude (Lat-Long) geographic coordinate system. 
The data structures of the Region Outlines were established based on the neighborhood's boundaries dataset made publicly available by the city's administration. A top-view rendering of the Outlines are presented in Figure~\ref{fig:regionOutlineFigure}. 


The \textbf{Connections} between Regions were established based on the real-world road network of the city. We obtained a dataset of the city's public transport systems, containing information on bus routes and stops. The dataset is also made publicly available by the city's administration and the information was adapted to our coordinate system. 

To represent the city with higher level of detail within LODUS, we extracted real-world geographic information from the OpenStreetMap (OSM) database. The dataset was restricted to the Lat-Long boundaries of the city. The base structure of the OSM database is a Node (OsmNode), which represents a unique point in space and is defined as a unique ID, a Lat-Long coordinate, a set of tags, and a set of version-control parameters. A Way (OsmWay) has the same definition combined with an ordered list of OsmNodes. An OsmWay can represent a variety of elements in the environment, including roads, building footprints, parks, bodies of water, and parking lots.

\textbf{Roads} can be distinguished from other elements due to their list representation. Lists containing the same OsmNode in the first and last index form a closed polygon, representing a \textbf{building} or area. Roads do not share that characteristic. Combined with that, a set of tags were considered when extracting information to represent the geometries.

\begin{itemize}
\item $building$: The Way represents a building footprint;
\item $height$: Indicates the height in meters of a building;
\item $floors$: Indicates the number of floors of a building. Each floor represents 3 meters;
\item $highway$: The Way represents a road;
\item $width$: Indicated the width in meters of a road; and
\item $lanes$: Indicates the number of lanes of a road. Each lane represents 3.7 meters;
\end{itemize}

\begin{figure}[!htb]
  \centering
  \subfigure[fig:topViewRoads][Primary Roads.]{\includegraphics[width=0.4\textwidth]{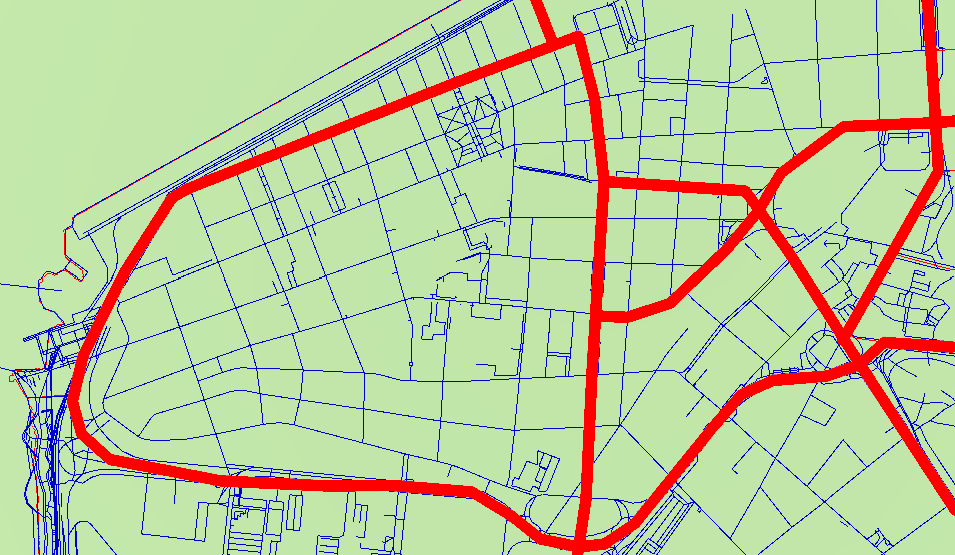}}
  \subfigure[fig:topviewRoadsBuildings][Building footprints.]{\includegraphics[width=0.4\textwidth]{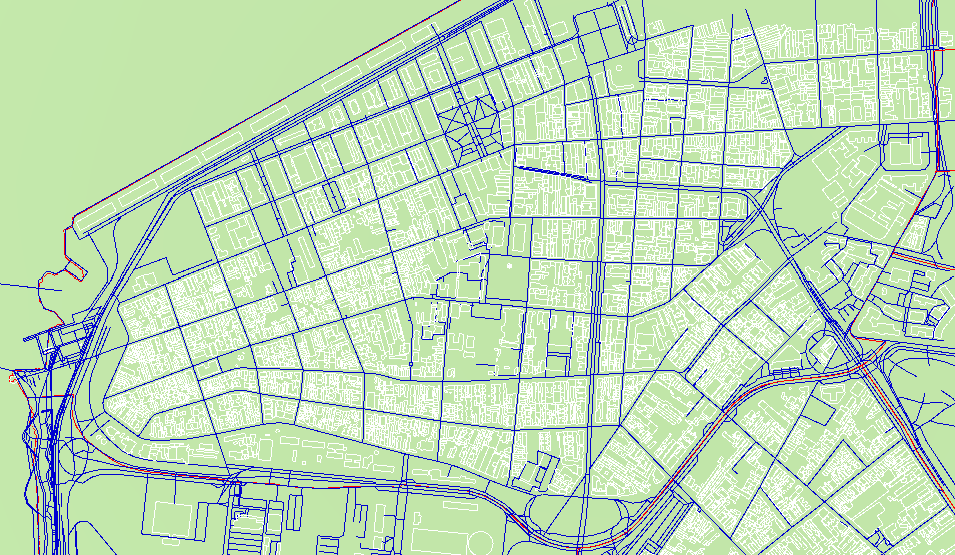}}
  \subfigure[fig:topViewMesh][Geometry creation.]{\includegraphics[width=0.4\textwidth]{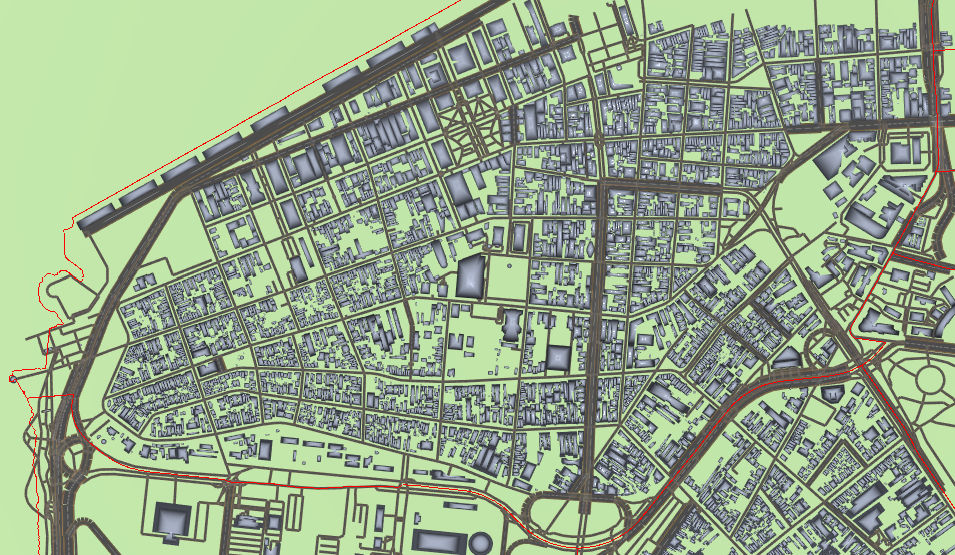}}
    \caption{Top view presenting part of the virtual city}
    \label{fig:topview}
\end{figure}

To represent the city layout in a 3-D environment, the Lat-Long coordinates of every element was modified through an Elliptical Mercator Projection algorithm. This method was selected over the Spherical Projection (also referenced as Web Mercator) due to a higher accuracy on aspect ratios and angles for objects anywhere on Earth. 
Figure~\ref{fig:topview} present the stages of a 3-D representation of part of the city. First, all road segments were loaded in the environment. Next, we defined a small set of primary roads, remaining segments were classified as secondary roads. Then, our Cycle Detection Algorithm was applied to compose the City Blocks. 

Building footprints were then loaded within the identified Blocks. Finally, a simple 3-D geometry was created for every element. Road segments are represented through an oriented quadrilateral connecting each pair of points with widths defined by the overall road. Buildings were generated via extrusion of their footprints. 


\subsection{Populating the City}

With the environment coherently created, LODUS also allow us to simulate the population behavior applying a multi-level approach. We are able to specify and visualize macroscopic simulation (e.g, the motion flow of clouds of people moving among neighbourhoods)  as well as microscopic simulation inside the buildings at the deeper level of abstraction.

To visualize the macroscopic flow of population, we feed a BioClouds simulation with city data given the following parameter setting:
\begin{enumerate}\itemsep=0pt
\item marker cell area of $16\,sqm$;
\item desired cloud speed of $10\,m/s$;
\item desired cloud density of $1\,agent/sqm$;
\item each region has its population broken up into clouds of $1000$ agents, and
\item cloud goals are randomized, from the list of region centers, whenever a cloud reaches its current goal.
\end{enumerate}
We use a large cloud speed value, $10\,m/s$, to account for urban mobility options, such as cars, public transportation, etc.

\begin{figure} [htb]
    \centering
    \includegraphics[width=0.3\textwidth, trim={6cm, 0cm, 6cm, 0cm}, clip]{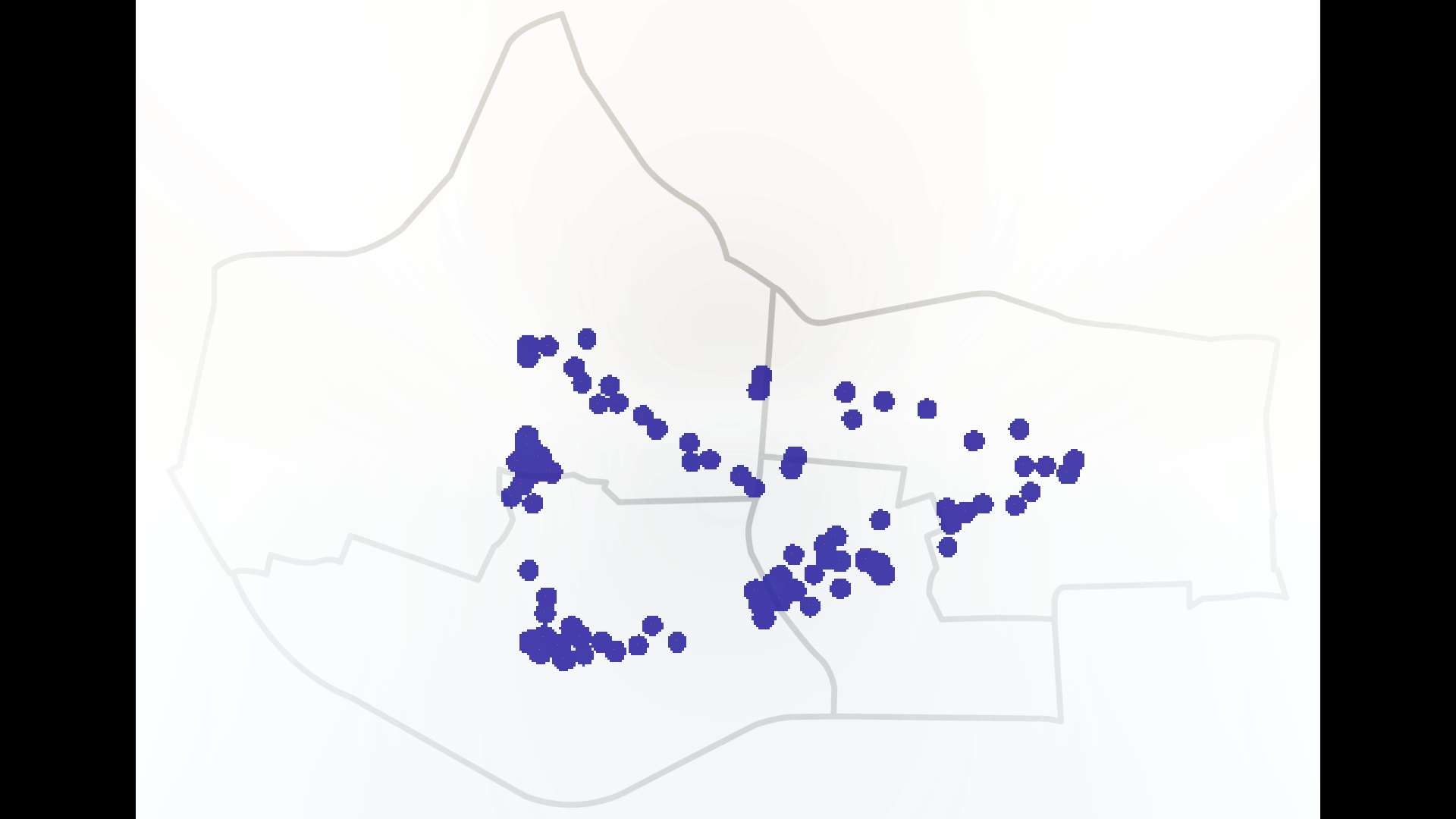}
    \caption{The population of Figure~\ref{fig:regionOutlines}, modeled with BioClouds.}
    \label{fig:bioclouds}
\end{figure}

Figure~\ref{fig:bioclouds} shows a BioClouds simulation of the population flow in the environment, as described in Figure~\ref{fig:regionOutlines} and the above parameters.

\subsection{Epidemic Contagion}

\begin{figure}
    \centering
    \includegraphics[scale = 0.28]{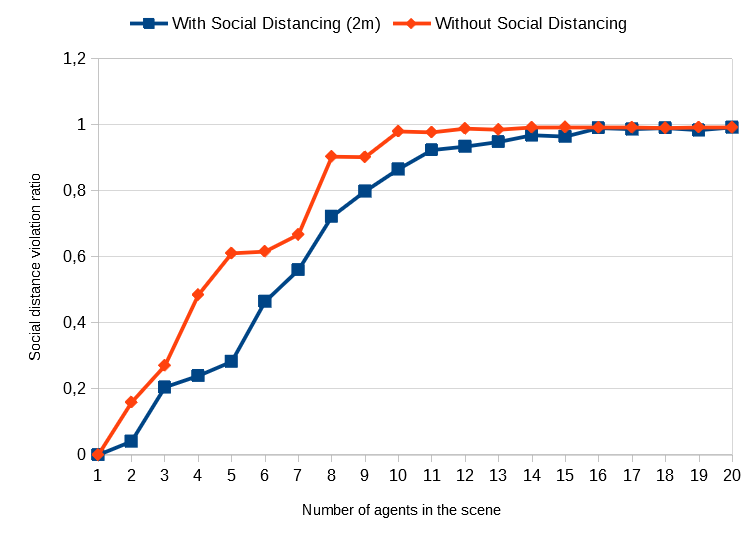}
    \caption{Contagion rate given the number of agents in the scenario (x-axis). This ratio is estimated by computing the number of simulation frames in violation of social distance over the total number of frames.}
    \label{fig:contagionRatio}
\end{figure}

After simulating the proposed scenario varying the number of agents in the scene, we extracted data about social distancing violation. Figure~\ref{fig:contagionRatio} depicts the results of simulating $2$ to $20$ agents in a scenario of $352m^2$. This results in densities of $0.00568 agents/m^2$ to $0.0568 agents/m^2$. We also compare the results obtained when agents are not programmed to attempt keeping social distancing. In the graphic presented in Figure~\ref{fig:contagionRatio}, when simulated with only two agents, there is already a considerable difference comparing with and without social distancing. Also, as the number of agents rise, the curve without social distancing rises faster, denoting that social distancing can indeed diminish contagion hazard. 
On the other hand, the curve with social distancing also rises fast, showing that indoor places may became hazardous places when overcrowded, making it impossible to keep social distancing, even if we try. This result emphasizes the needs to keep indoor numbers low, aiming for preserving public health.

Whenever the macroscopic simulation requires a contagion event, a random number is generated in the range $[0,1]$. If this number is grater than the contagion rate, consulting the curve on Figure~\ref{fig:contagionRatio} and interpolating the points, the input population is submitted to the SIR dynamics. The parameters used are $\beta = 0.7$ and $\gamma = 0.35$, as described in Section~\ref{sec:populationAndContagion}. After this process, individuals of the population move from Susceptible state to Infected, and from Infected state to Removed. The number of recovered and deceased is estimated based on statistics for COVID-19 and profile of individuals, such as age, applied to the recovered group.

\section{Final Remarks}

This paper presented LODUS, a framework for urban simulations applying different levels-of-detail. LODUS may be a decision support tool collaborating with city managers work. Obtained results have shown the process to rebuild and simulate a real city on different levels of abstraction. In addition, specific aspects were considered, simulating even social distance, an important aspect on pandemic period.

Furthermore, presented results highlight the main contributions of the work: \textit{i)} To reproduce, into a 3D environment, a real city according real geographical and population data; \textit{ii)} to provide a module to simulate important points on a pandemic period, such as contagion and social distance; and \textit{iii) }to develop an application able to be used by governments and other city managers to study and predict attention points in the city.

The field of smart cities is a huge area with several open research points. Beyond that, we expect to identify many potential future work on LODUS.

\section*{Acknowledgment}

The authors would like to thank the Brazilian Research Agencies FAPERGS, CAPES and CNPQ.



%


\bibliographystyle{IEEEtran}
\bibliography{references.bib, SAGabriel.bib}


\end{document}